\documentclass[numbers]{article}
\usepackage[numbers]{natbib}
\usepackage{multirow}

\usepackage[preprint]{neurips_2024}

\usepackage{tabularx}
\usepackage[utf8]{inputenc} 
\usepackage[T1]{fontenc}    
\usepackage{hyperref}       
\usepackage{url}            
\usepackage{booktabs}       
\usepackage{amsfonts}       
\usepackage{nicefrac}       
\usepackage{microtype}      
\usepackage{xcolor}         
\usepackage{algorithm}
\usepackage{algpseudocode}
\usepackage{graphicx}

\title{Tree Search for Simultaneous Move Games via Equilibrium Approximation}

\author{%
  Ryan Yu \\
  Department of Computer Science\\
  Boston University\\
  Boston, MA, 02215 \\
  \texttt{ryu1@bu.edu} \\
  \And
  Alex Olshevsky\\
  Department of Electrical and Computer Engineering \\
  Boston, MA, 02215\\
  alexols@bu.edu \\
  \And
  Peter Chin \\
  Department of Engineering \\
  Hanover, NH 03755\\
  pc@dartmouth.edu \\
}

\begin{document}

\maketitle

\begin{abstract}






Neural network supported tree-search has shown strong results in a variety of perfect information multi-agent tasks. However, the performance of these methods on partial information games has generally been below competing approaches. Here we study the class of simultaneous-move games, which are a subclass of partial information games which are most similar to perfect information games: both agents know the game state  with the exception of the opponent's move, which is revealed only after each agent makes its own move. Simultaneous move games  include popular benchmarks such as Google Research Football and Starcraft.

In this study we answer the question: can we take  tree search algorithms trained through self-play from perfect information settings and adapt them to  simultaneous move games without significant loss of performance? We answer this question by deriving a practical method that attempts to approximate a coarse correlated equilibrium as a subroutine within a tree search. Our algorithm works on cooperative, competitive, and mixed tasks. Our results are better than the current best MARL algorithms on a wide range of accepted baseline environments.  
\end{abstract}

\section{Introduction}
Multi-agent reinforcement learning (MARL) algorithms train multiple agents interacting in a shared environment. The challenge is that, for each agent, the actions of the other agents are influencing the received rewards; the evolution of other agents policies during training makes the environment appear non-stationary to each agent. 
While reinforcement learning methods have achieved tremendous successes for algorithms and tasks where agents are fully cooperative \cite{zhang2021multiagent}, successes in cooperative settings with partial information have been more muted. 


Here we study the class of simultaneous-move games, which are a subclass of partial information games which are most similar to perfect information games: both agents know the game state  with the exception of the opponent's move, which is revealed only after each agent makes its own move.  These are partial information games most similar to full information games. While cooperative simultaneous-move tasks can be easily equated to maximizing the rewards over a joint policy, training agents in competitive simultaneous-move tasks is challenging. State of the art results often use human injected knowledge, either in the form of a human-designed fixed opponent against which one can train \cite{wei2017}, or in the form of imitation learning of human data \cite{Vinyals2019GrandmasterLI}. This to be contrasted with Markov Chain Tree Search (MCTS) which can learn purely from self-play. 

Here we focus on simultaneous move games (which can be either cooperative or competitive) with discrete actions and finite state space. We are interested in training agents to learn from self-play alone. Our main contribution is a new  MARL algorithm which works by modifying MCTS to approximate a coarse correlated equilibrium within the tree search. We show that our algorithm consistently and significantly outperforms competitive algorithms.

\section{Background Information}


We define an $N$-player stochastic game (SG) as ($S, H, \{A_i\}_{i \in N}, T, \{U_i\}_{i \in N}, \gamma$), where  $\mathbf{S}$ is the set of all states shared by all $\mathbf{N}$ players, $\mathbf{H}$ is the horizon (the maximum number of time steps),  $\mathbf{A_i}$ is the action space for player $i$ yielding the decomposition $\mathbf{A} := \mathbf{A_1} \times \dots \times \mathbf{A_N}$, $\mathbf{T}: (S \times A) \rightarrow S'$ is the state transition function, $\mathbf{U_i}: (S \times A) \rightarrow \mathcal{R}$ is the utility function for each player $i \in N$, and finally $\gamma$ is the discount factor. Note that $\mathbf{T}$ is defined on $S \times \mathbf{A}$, i.e., the next state is a function of  the actions taken by all the players. Finally, we denote  by $\Delta(S)$ a distribution over the starting states. 


\subsection{Deep MARL Training}
Our goal is to train a set of agents, defined by their policies, $\{\pi_i\}_{i \in N}$, where each policy, $\pi_i: S \rightarrow A$. Here the simultaneous-move nature of the game will come through, as each policy $\pi$ takes only the state (not the actions of the other players that are unknown) as arguments.  We assume a finite set of actions. 

Training through deep reinforcement learning is usually comprised of two iterating steps: data generation and network training. Data generation aims to create a data set, $\mathcal{D}_t$, that the NN samples from to train. A NN is used to approximate the value function and policy function, given the following loss functions $$L(\theta_t) = \frac{1}{|\mathcal{D}_t|}E_{s \sim \mathcal{D}_t}L(g_t(s),\hat{g_{\theta_t}}(s))$$

where $g_t$ is the value or policy function at time step $t$, $\hat{g_{\theta_t}}$ is the value or policy prediction of the network at time step $t$ , $L$ is an appropriate loss function, and $\theta_t$ represents the network parameters at time step $t$. 

The data generation is usually accomplished through repeated interaction with the stochastic game. The recorded interactions are then used to train the same NNs \cite{Lee2021}. This is done simultaneously: the NN influences the data it receives at the next iteration. 


\subsection{Challenges of MARL}
We briefly summarize the difficulty of deep MARL training: (1) \textbf{non-stationary environments}, (2) \textbf{learning objectives}, (3) \textbf{curse of dimensionality}, and (4) \textbf{scaling beyond two players}. Empirically successful algorithms address at least one of these issues \cite{yang2020}. 

To begin with, (1) the environment appears \textbf{non-stationary} to a single agent $i$ if the policies of the other agents change during training. Of course, the environment is not actually non-stationary; but each player only observes a sequence of actions and rewards, which is the standard RL setup, and from this perspective the environment can appear non-stationary to that agent. 

Second, setting the (2) \textbf{learning objectives} of agents to maximize their own reward leads to sub optimal policies if agents have conflicting reward functions.

Next, (3) \textbf{curse of dimensionality} refers to the exponential growth of the number of joint actions with respect to the number of players. The size of the joint action space is $|A|^N$, assuming that all players share the same action space. 

All of these reasons become more potent as we (4) \textbf{scale beyond two players}. The curse of dimensionality gets worse because $N$ is in the exponent, there are more subtle ways for objectives to conflict, and the environment may appear to change more significantly if more agents are continuallyadopting their policies. 

Previous works are bottle-necked by challenge (2). This study provides empirical results on a novel algorithm that directly addresses (2) as well as touches upon (1).  

The four challenges are closely linked to one another. This study addresses all of them. We use a game theoretic based no-regret learning to provide a (1) strong learning objective and address (2) non-stationary environments, and we use deep learning function approximation to mitigate the (3) curse of dimensionality and (4) provide the ability to scale beyond two players without heavy penalty in terms of time and resources. 

\subsection{Game Theory and online learning}
There is a close relationship between multi-agent learning and game theory which we will exploit in this work. 

The concept of an equilibrium provides a strong learning objective in multi-agent settings. The most popular equilibrium, the Nash Equilibrium, describes a set of strategies in a two player zero-sum (2p0s) game in which neither player gains any benefit from changing strategies. While computing a NE is ideal, it was shown to be PPAD-complete even in 2p0s games \cite{NisaRougTardVazi07}.

Less restrictive forms of equilibrium can be approximated using no-regret learning. In this study, we will attempt to approximate an $\epsilon$-coarse correlated equilibrium (CCE). A CCE, $\sigma$, is defined as $$\forall i, a_i' \ \ \ E_{s \sim \sigma} \ c_i(a)\leq E_{s \sim \sigma} \ c_i(a_i', a_{-i}) + \epsilon$$ where $i$ represents a player, $a_i'$ represents an action different from the recommended action, $a$, and $c_i$ represents the cost of following a strategy. At first glance, this might look the normal definitions of mixed strategy Nash equilibrium, but observe that the joint state $s$ is being sampled from some distribution $\sigma$. Thus this definition says that there exists a joint distribution of the strategies such that deviations do not benefit each player, provided that all the remaining players sample from the same correlated distribution.

If all players in a SM game use no-regret online learning, then their time-averaged policies converge to the set of CCEs \cite{TardosCCELecture, RoughgardenLectures}. We define the regret at time step T as $$R_T = \max_{i \in [K]} \mathbb{E} [\sum_{t=1}^T l_{t,I_t} - \sum_{t=1}^T l_{t,i}]$$ where the player has an action space of size $K$, and $l_{t,k}$ represents the loss experienced at time step $t$ for action $k\in K$. It may seem puzzling that individual actions by each agent made independently converge to a correlated action distribution, but note that agents are effectively responding to each other so that correlation can be introduced into their time-averaged policies. 

No-regret learning measures the difference in loss compared to the best single action in hindsight. Successful learning in this framework provides guarantees that the regret grows sub-linearly with respect to $T$ in expectation or with high probability. Below, we will utilize no-regret learning algorithms, EXP-IX \cite{Neu2015ExploreNM} and EXP-WIX \cite{kocak16} which are known to have the property of no regret-learning with high probability.

\section{Related Work}
All MARL algorithms fall between two extremes: decentralized methods and centralized methods. Decentralized methods train agents simultaneously but independently. Independent Q-learning (IQL) \cite{tampuu2017} and independent proximal policy optimization (IPPO) \cite{dewitt2020independent} are primary examples of decentralized algorithms. Each agent repeatedly improves their Q-value approximations, in the case of IQL, or their policy, in the case of IPPO, through repeated interactions with the environment. 

On the other hand, centralized methods learn a policy over the joint action space, but are largely restricted to environments where agents share reward functions. A middle ground between the two extremes are algorithms with centralized learning and decentralized execution (CTDE).

For instance, multi-agent proximal policy optimization \cite{Yu2021TheSE} (MAPPO) is an example. MA-PPO demonstrated superior performance to other popular CTDE MARL algorithms such as Simplified Action Decoder (SAD), Value Decomposition (VDN) and QMIX \cite{rashi2018} plus its variants on several benchmarks. These algorithms are restricted to  cooperative environments.



Several studies have attempted an obvious expansion from purely cooperative to cooperative-competitive environments: behave selfishly. Agents that behave selfishly (i.e. do not consider the policies or actions of other players) will encounter identical states with different value estimates and will have difficulty learning in mixed cooperative and competitive environments\cite{tampuu2017,zawadzki2014,Lee2021}). 

Another method for branching into competitive environments is to freeze the policies of certain agents during training. This way, the environment becomes stationary with respect to a single agent \cite{Vinyals2019GrandmasterLI}. A common implementation of this concept is neural fictitious self-play \cite{Heinrich2016}, where an agent plays against frozen past iterations of themselves and the pool of past policies grows during training. 

Next, there is a variation of policy freezing where agents either have explicit access to, or maintain their own approximation of, other agent policies. A popular example of such an algorithm is Multi Agent Deep Deterministic Policy Gradient (MADDPG)\cite{Lowe2017}. This directly addresses the problem of non-stationary and allows for training on cooperative, competitive, and mixed environments. We focus our performance comparisons against MADDPG as it has the most similar motivations and applications as this study. 

The final approach some studies take to address challenges (1) and (2) is to attempt to approximate an equilibrium. Counterfactual regret minimization \cite{Zinkevich2008, Neller2013AnIT} provides a powerful algorithm for approximating a Nash-equilibrium in 2p0s tasks. The algorithm aims to minimize regret. 
\cite{Zinkevich2008, Neller2013AnIT} demonstrate that by attempting to minimize the regret, they also minimize the exploitability of their policy thus approximating a Nash-equilibrium.  Policy Space Response Oracles (PSRO) \cite{lanctot2017} addresses poor convergence in multi-agent settings due to other agents' policies. 
At each iteration, a new policy is added that approximates the best response so the meta-strategy of the other players. It has several variations such as joint PSRO (jPSRO) that are improvements upon the base algorithm. The limitation for equilibrium approximation algorithms is that they are not easily applied to tasks with larger state and action spaces. The majority of testing regarding such algorithms have been on small tasks.

\section{Methodology}

Our approach is based on two core ideas. As already implicit above, the first idea is to train a separate neural network $\pi_i$ for each agent: this potentially gets rid of the the curse of dimensionality. Indeed, although there are $|A|^n$ joint actions, by training $\pi_i: S \rightarrow \mathbf{A}_i$ for all $i=1, \ldots, n$, we avoid an exponential blowup of the action space. However, there is a pretty large trade off  in that we have effectively made agent decisions independent of each other -- which is clearly highly suboptimal. Indeed, agents could potentially gain from correlating their actions, which they cannot do under this strategy. 

Out second idea is to mitigate the loss from this by using update rules  which approximate a coarse correlated equilibrium (CCE); we view this as a "second best" solution to making correlated decisions across agents. Recall that a CCE is similar to a standard mixed Nash equilibrium, except that there is a {\em joint} distribution taken by all the agents which makes  deviations gainless. By training agent policies to learn a CCE, we effectively bypass a major limitation of training independent policies. The idea is that even though the individual decisions are made separately, the agents are replying to each other, so that, in the limit, their time-averaged policies converge to a correlated action profile which is good in the sense of being a CCE. 

This leads to the question of how to build dynamics that attain CCE in our setting. We build on recent work \cite{daskalakis2022complexity} which shows that EXP-IX, a standard algorithm in online learning with asymptotically vanishing regret, can learn CCEs. We thus replace the standard value estimation methods in MCTS based on UCB estimates with EXP-IV based estimates. Details are given below. 

We call our algorithm NN-CCE. It is trained by iterating through three main steps: (1) gathering trajectories $D$, (2) processing trajectory information, (3) training a model on $D$.  

We first gather $K$ trajectories in a given environment, each of length $H$. Next, we process our trajectories in reverse  order. For the set of states in time step $H-1$ \textbf{over all trajectories}, we train a Q-value network,  $Q_{H-1}: s_{H-1} \times A \rightarrow \mathbb{R}$ to output $Q$-value estimates of state-action pairs. Using $Q_{H-1}$, we perform Multi-Agent EXP-IX to approximate a CCE policy and value estimate for every state over all trajectories at time step $H-1$, $\{v(s_{H-1})\}_i$ and $\{p(s_{H-1})\}_i \ \forall i \in [K]$ (Algorithm \ref{alg:MAEXP}). 

After, we use $\{v(s_{H-1})\}_i \forall i \in [K]$, to train a new Q-value network $Q_{H-2}: s_{H-2} \times A \rightarrow v(s_{H-1}) \in \mathbb{R}$, and repeat CCE approximation and Q-value training until time step $h=0$. In total, we would train $H$ Q-value networks per agent. Each Q-network takes as data the trajectory info for its respective time-step. 

Finally, we train a policy model on all CCE policies calculated during our trajectory processing phase to form a final model that outputs a policy distribution. 

A detailed algorithm is provided in Algorithms \ref{alg:AlgorithmOverview}, \ref{alg:GenerateDataset}, and \ref{alg:MAEXP}. 

\begin{algorithm}[tb]
\caption{NN-CCE approximation. This is a standard on-policy value-based method except we train a value network for each time step in reversed time step order.}
\label{alg:AlgorithmOverview}
\textbf{Input}: $G$, Stochastic Game\\
\textbf{Input}: $H$, finite horizon \\
\textbf{Input}: $K$, number of trajectories \\
\textbf{Input}: $M_H$, Q-value network\\
\textbf{Input}: $\pi_0$, Initial policy network \\
\begin{algorithmic}[1] 
\State $R \gets \{\emptyset\} $ 
\For{$t = 1 , 2, \dots, T$}
    \For{$i \in K$}
        \State $D_i \gets $ GenerateDataset($\pi_i, G, K$) 
        \State Let $b_{h'} \in D_i$ represent the set of all states  from time-step $h'$ for player $i$
    \EndFor
    \For{$h = H-1, H-2, \dots, 0$}
        \State $v_h, \pi_h \gets$ MA-EXP-IX$(s, M_{h+1}) \ \forall s \in b_h$
        \State $M_h \gets$ TrainValue($\{v_{h'} : h' = h\}$)
    \EndFor
    \State $\pi_t \gets $ TrainPolicy($\{\pi_h\}$)
\EndFor
\State return $R$
\end{algorithmic}
\end{algorithm}

\begin{algorithm}[tb]
\caption{GenerateDataset}
\label{alg:GenerateDataset}
\textbf{Input}: $G$, Stochastic Game\\
\textbf{Input}: $H$, finite horizon \\
\textbf{Input}: $K$, number of trajectories \\
\textbf{Input}: $\pi$, policy network \\
\begin{algorithmic}[1] 
\State $\{b_h \gets \{\} \}_{h \in H}$
\For{$k = 1 , 2, \dots, K$}
    \State $s_1 \gets \Delta(S)$ \Comment{sample starting state}
    \For{$h = 1, 2, \dots, H$}
        \State $b_h \gets b_h \cup s_h$
        \State $a_h \gets \pi(s_h)$
        \State $s_h \gets T(s_h, a)$ \Comment{$T$ represents the transition function}
    \EndFor
\EndFor
\State return $\{b_h \gets \{\} \}_{h \in H}$
\end{algorithmic}
\end{algorithm}

\begin{algorithm}[tb]
\caption{MA-EXP-IX. This is a standard EXP-IX algorithm from \cite{Neu2015ExploreNM} except we provide additional details because multiple players are all simultaneously using EXP-IX no-regret learning.}
\label{alg:MAEXP}
\textbf{Input}: $A$, Number of actions \\
\textbf{Input}: $T$, max time step \\
\textbf{Input}: $N$, number of players \\
\textbf{Input}: $Q$, Q-value estimation network \\
\textbf{Input}: $h$, current time horizon \\
\textbf{Input}: $H$, max time horizon \\
\textbf{Input}: $s$, state \\
\textbf{Output}: weight matrix, $w$ and value estimate, $v$
\begin{algorithmic}[1] 
\State $w \gets \overrightarrow{1} \in \mathbb{R}^{N \times A}$ 
\State $v \gets \overrightarrow{0} \in \mathbb{R}^{N}$
\For{$t = 1, 2, \dots, T$}
    \State $j \gets \{\}$
    \For{$n = 1, 2, \dots, N$}
        \State $p_{t,i,n} = \frac{w_{t,i,n}}{\sum_{j=1}^{A}w_{t,j,n}}$
        \State Draw $I_{t,n} \sim p_{t,n} = (p_{t,1,n},p_{t,2,n},\dots,p_{t,A,n})$
        \State $j \gets j \cup \{I_{t,n}\}$
    \EndFor
    \If{$h = H$}
        \State Observe loss $l_{t,j} = (l_{t,I_{t,1}}, \dots ,l_{t,I_{t,n}})$
    \Else
        \State $l_{t,j} \gets Q(s,j)$ 
    \EndIf
    
    \For{$n = 1, 2, \dots, N$}
        \State $v_n \gets v_n + l_{t,_{t,n}}$
        \State $\widetilde{l_{t,i,n}} \gets \frac{l_{t,i,n}}{p_{t,i,n}+\gamma}\mathbb{I}_{\{I_t=i\}}$ for all $i \in [A]$
        \State $w_{t+1,i,n} \gets w_{t,i,n}e^{-\eta \widetilde{l_{t,i,n}}}$ for all $i \in [A]$
    \EndFor
\EndFor
\State return $w$ and $v$
\end{algorithmic}
\end{algorithm}

\subsection{Experimental Parameters}
All Q-value and policy network parameters are given below. Let us define $I$ as the size of the input, $J$ as the size of the joint action space, $P$ as the size of the policy space, $H$ as the finite horizon, and $N$ as the number of players.  All networks are trained using an Adam optimizer with learning rate $5e-5$. 
\begin{center}
\begin{tabular}{ |p{2cm}||p{3cm}|p{2cm}|p{2cm}| p{1cm} |}
 \hline
 \multicolumn{5}{|c|}{Q-value Network Parameters} \\
  \hline
 Sub-Network Name & Architecture & Learning Rate & L2-Regularization & Dropout\\
 \hline
 Representation Network & [$I$, 256, 256, 32] & 5e-5 & 1e-4 & 0.5\\
 \hline
 Q-value prediction Network & [32 + $J$, 256, 256, S] & 5e-5 & 1e-4 & 0.5 \\
 \hline
\end{tabular}
\end{center}

\begin{center}
\begin{tabular}{ |p{2cm}||p{3cm}|p{2cm}|p{2cm}| p{1cm} |}
 \hline
 \multicolumn{5}{|c|}{Policy Network Parameters} \\
  \hline
 Sub-Network Name & Architecture & Learning Rate & L2-Regularization & Dropout\\
 \hline
 Representation Network & [$I$, 1028, 1028, 64] & 5e-5 & 2e-4 & 0.6\\
 \hline
 Q-value prediction Network & [64, 1028, 1028, $P$] & 5e-5 & 2e-4 & 0.6 \\

 \hline
\end{tabular}
\end{center}

For a given environment, we trained a total of $H*N$ Q-value networks, one for each player and time step. The number of Q-value networks could be reduced if the game was fully competitive or fully cooperative. In this type of environment, only $H$ Q-value networks were trained. For every environment, a total of $N$ policy networks were trained. 

\section{Points of Comparison}

\subsection{Environments}
In this study we focus on 4 main environments, where each environment contains between 2-6 unique scenarios. We define a scenario as a unique SG within an environment. Each environment was chosen because it is an open-source widely used MARL library with optimized performance to allow for fast training and was used by at least three other popular algorithms. Of all environments that fit this description, these four were the most well cited and used. 

\textbf{OpenSpiel} \cite{LanctotOpenSpiel2019} . A collection of $n$-player imperfect information games. Scenarios from this environment were small enough such that equilibrium approximation methods could converge onto a solution in a reasonable amount of time. Two scenarios are used: Goofspiel-6 (6-card variant) and Laser Tag. 

\textbf{Google Football Research} \cite{kurach2020google}. A team based mixed cooperative competitive football simulation environment. For this study we use smaller scale environments rather than a full game. We train agents to play as both teams and do not used fixed algorithms in our training process. Six scenarios are used: 3v1, CA(easy), CA(hard), Corner, PS, and RPS. 

\textbf{Multi Agent Particle Environment} \cite{Lowe2017,mordatch2017emergence}. A multi-agent particle environment that has a mix of cooperative, competitive, and cooperative-competitive tasks. We focus on three competitive and cooperative-competitive scenarios: Adv, Tag, and Push. 

\textbf{Starcraft Multi-agent Challenge} \cite{Samvelyan2019SMAC} A multi-agent version of the popular real-time strategy game Starcraft. In this variant, all pieces on a single team are controlled simultaneously at each time step. Three scenarios are used: 3s vs 3z, 3s vs 4z, 5m vs 6m. 

\subsection{Compared Algorithms and Evaluation Metrics}
We divide the set of algorithms we compared against into three main groups. We chose each of the algorithms tested because it is or was a recent state of the art algorithm for a respective environment, or it is a  popular algorithm that serves as a useful benchmark in assessing our algorithm. 

In addition, all algorithms we compared against are easily accessible with open-source code as well as a detailed list of training parameters that facilitate robust replication of their results. 

\textbf{Equilibrium Approximation Algorithms}: NN-CCE (ours) compared against PSRO, JPSRO, CFR. Each of these three algorithms are not meant to scale to larger environments. Their application is limited to smaller competitive tasks from the OpenSpiel environment. For a comparison metric, we compared a direct head-to-head win rate of our algorithm versus the opposing algorithm for each scenario in OpenSpiel. All scenarios from open-spiel were symmetric 2p0s. In a single trajectory we labeled recorded the total points accumulated by our agent and the contemporary agent. The winner was determined as the side that accumulated more points. All algorithms in the comparison were trained and assessed over 10 random seeds. 

\textbf{Cooperative Algorithms}: NN-CCE (ours) compared against MAPPO, MADDPG. Each of these algorithms are able to be applied to purely cooperative tasks within GFR and SMAC. They are both popular algorithms with tested open-source implementations. In addition they both demonstrate superior performance against a wide arrange of other contemporary algorithms \cite{Yu2021TheSE,Lowe2017}. For a comparison metric, we compared total accumulated score in testing scenarios of our algorithm to the competitor. 

\textbf{Competitive Algorithms}: NN-CCE (ours) compared against MADDPG, Simultaneous Move MCTS. All three of these algorithms can be applied to larger scale competitive scenarios within MPE, GFR and SCMAC. MADDPG has a tested open-source implementation. We implemented Simultaneous Move MCTS locally, and its detailed are found in the appendix. For a comparison metric, we compared a direct head-to-head win rate of our algorithm compared to the opposing algorithm for each scenario in the three environments listed. All competitive scenarios from MPE, GFR, and SMAC were asymmetric and two team-based. For all comparisons, our algorithm and a contemporary agent played as 

\section{Performance Results}
\begin{table}[t]
\centering
\parbox{0.49\linewidth}{
    \resizebox{0.49\columnwidth}{!}{
        \begin{tabular}{|l|l|l|l|l|l|}
            \hline
            & NN-CCE & jPSRO & PSRO & CFR & R  \\
            \hline
            NN-CCE & - & 62\% & 70\% & 91\% & 100\%\\
            \hline
            jPSRO & 38\% & - & 63\% & 55\% & 85\%\\
            \hline
            PSRO & 30\% & 37\% & - & 52\% & 73\% \\
            \hline
            CFR & 9\% & 45\% & 48\% & - & 74\% \\
            \hline
            R & 0\% & 15\% & 27 \% & 26 \% & - \\
            \hline
        \end{tabular} 
        }
    }
    \parbox{0.49\linewidth}{
        \resizebox{0.49\columnwidth}{!}{
        \begin{tabular}{|l|l|l|l|}
            \hline
            & NN-CCE & C Q-Learning & JPSRO\\
            \hline
            NN-CCE & - & 100\% & 74\%\\
            \hline
            C Q-Learning & 0 \% & - & 7\%  \\
            \hline
            JPSRO & 26\% & 93\% & - \\
            \hline
            Random & 0\% & 36\% & 12\% \\
            \hline
        \end{tabular}
        }
    } \caption{Win Rate on ``Goofspiel-6" Scenario (Left) and  ``Laser Tag" Scenario (Right) from OpenSpiel. (Left) C Q-Learning results are not reported because it failed to beat the random opponent over many repeated trials. (Right) PSRO and CFR results are not reported because they failed to converge to a solution in a reasonable amount of time}
    \label{results:Openspiel}
\end{table}
   
\begin{table}[t]
\centering

\label{LaserTable}
\end{table}

\begin{table}[t]
\centering
\begin{tabular}{|l|l|l|l|}
    \hline
    & & MADDPG & S-MCTS \\
    \hline
    \multirow{3}{*}{MPE} & Adv & 82\% & 83 \%\\
    & Tag & 85\% & 90 \%\\
    & Push & 81\% & 87 \%\\
    \hline
    \multirow{3}{*}{GFR} & MA-PS & 60\%& 100 \%\\
    & MA-3v1 & 63\% & 99 \%\\
    & MA-C & 60\% & 100 \%\\
    \hline
    \multirow{3}{*}{SMAC} & 3s,vs,3z & 61\%& 100 \%\\
    & 3s,vs,4z & 64\% & 100 \%\\
    & 5m,vs,6m & 60\% & 100 \%\\
    \hline
\end{tabular}
\caption{Win rate on ``MPE" Tasks, Adv - Simple Adversary, Tag - Simple Tag Environment, Push - Simple Push. Win rate on ``Google Football Research" Tasks. PS - Pass and Shoot Scenario, 3v1 - Academy 3 v 1 with keeper Scenario, Counter - Counterattack Easy scenario. All scenario descriptions can be found in the Google Football Repository. The tag ``MA" refers to the multi-agent variant of the scenario, where multiple learning agents control all pieces, one agent per piece. Three tasks were chosen from the SMAC environment, which represent the number of pieces controlled by two players.}
\label{MPETable}
\end{table}

\textbf{Comparing to equilibrium approximation algorithms}. First, we apply our algorithm to the relatively small scenarios of Goofspiel and Laser Tag from the Openspiel environment. These results are summarized in Tables \ref{results:Openspiel}. As can be seen, NN-CCE has a higher win rate across all three tasks compared to other equilibrium approximation algorithms. All algorithms we compared against do not scale to larger environments. Therefore, we compared them in the smaller scaled scenarios of OpenSpiel. 


\textbf{Comparisons in competitive scenarios}. In the next set of experiments, we assessed NN-CCE on tasks that involved controlling multiple pieces in competitive environments: MPE GFR, and SMAC. In our assessment we compared against a popular multi-agent algorithm Multi-Agent MADDPG. The results of these experiments are summarized in Tables \ref{MPETable}. Across all three environments (MPE, GFR, and SMAC), our algorithm had a higher win rate compared to MADDPG and SM-MCTS.

\begin{figure}[!tbp]
  \centering
    \includegraphics[width=0.5\textwidth]{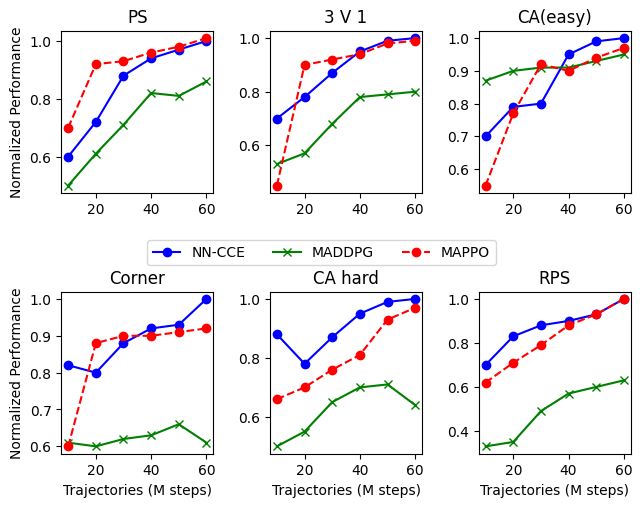}
    \caption{Results on GFR. NN-CCE (ours) and MADDPG are trained via self-play, MAPPO is trained against a fixed algorithm opponent.}
  \hfill
  \label{results:GraphsGFRandSMAC}
\end{figure}

\textbf{Comparisons in cooperative scenarios}. In the final set of experiments we compare our method against popular cooperative algorithms over GFR and SMAC environments. We can see in table \ref{PPOGFR} that that our method, NN-CCE, demonstrates marginal to high success-rate improvement over MA-PPO across 6 different scenarios in GR. {\em It is important to note that MA-PPO trains against a fixed opponent in scenarios where an opponent is present, such as scenario 3v.1, where as our agent trains against an adaptive policy (itself) in such a case.} Figure \ref{results:GraphsGFRandSMAC} also visualize the performance as a function of trajectories learned by our algorithm, MADDPG, and MAPPO. 

\begin{table}[t]
\centering
\begin{tabular}{|l|l|l|l|}
    \hline
    Scen. & \textbf{NN-CCE} & MA-PPO & S-MCTS \\
    \hline
    3v.1 & \textbf{$89.00_{(1.50)}$} & $88.03_{(1.06)}$ & $65.01{(2.21)}$ \\
    \hline
    CA(easy) & \textbf{$90.03_{(1.76)}$}  & $87.76_{(1.34)}$ & $80.02_{(2.03)}$ \\
    \hline
    CA(hard) & \textbf{$79.03_{(5.85)}$} & $77.38_{(4.81)}$ & $55.15_{(1.22)}$ \\
    \hline
    Corner & \textbf{$70.03_{(1.03)}$} & $65.53_{(2.19)}$ & $44.19_{(1.77)}$ \\
    \hline
    PS & \textbf{$94.2_{(1.06)}$} & $94.92_{(0.68)}$ & $78.09_{(1.23)}$\\
    \hline
    RPS & \textbf{$75.8_{(1.99)}$} & $76.83_{(1.81)}$ & $65.55_{(0.50)}$ \\
    \hline
\end{tabular}
\caption{Success rate comparison between NN-CCE and MAPPO on different scenarios within the GFR environment. Results for MAPPO are taken from \cite{Yu2021TheSE} Average and standard deviation success rates are reported over six random seeds for each scenario. S-MCTS results are based on our own implementation}
\label{PPOGFR}
\end{table}

\section{Conclusion and Future Work}
In this study our aim was to create an algorithm that could perform well in multi-agent scenarios and trained through self-play, with little human-knowledge injection. To accomplish this we proposed a novel method to use NN to estimate a CCE for any given task. We demonstrate that our algorithm obtains higher performance against its contemporaries from both game theory and deep MARL across a variety of benchmarks. In addition, we show that our agent has higher consistency in agent performance over many agents, and can achieve the same if not better performance than its contemporaries using at most half of the sampled trajectories. 

We demonstrated an improvement against other equilibrium estimation algorithms (PSRO, CFR) for smaller tasks, and and improvement against the current state of the art (MADDPG) for cooperative competitive environments in larger tasks. In conjunction, we also demonstrate a higher empirical consistency factor to our algorithm compared to MADDPG; our algorithm is much more likely to show improvement from baseline across MPE and GFR tasks. Lastly, we demonstrate that our algorithm also shows improvement over contemporary multi-agent algorithm MAPPO in purely cooperative tasks. 

The algorithm addresses two shortcoming of current MARL algorithms. Firstly, it can adapt to environments where agents with competing objectives exists. There is a small pool of algorithms that can successfully work in mixed cooperative competitive environments; of this pool our algorithm, NN-CCE, demonstrated higher performance across a variety of tasks. Secondly, it further further detaches itself from the need of human injected knowledge (typically used in the form of a human-designed agent to train against) and can therefore be used in environments where a strong fixed policy is not well known. 

\textbf{Limitations}.
The main limitation of this work is that training process is longer than other deep-MARL algorithms. This elongation stems from the fact that we now train a Q-value network for each time step, and thus need to process $K$ data points using MA-EXP-IX (Algorithm \ref{alg:MAEXP}) at every time step. However, the trade-off we make for longer training is faster execution. Our time-cost at execution is as fast as any other deep-MARL algorithm: an observation is passed to the policy network and a policy distribution is output.


\textbf{Appendix Information}. The appendix is divided into two main sections. Section 8.1 describes ablation studies that were influential to NN-CCE approximation performance. Section 8.2 Describes our own implementation of simultaneous-move MCTS. 

\textbf{Future work}. The clear drawback of our method is its limitation to discrete data tasks. Although we have higher performance than MADDPG in this realm, MADDPG boasts the ability to work in tasks with continuous space and continuous action spaces. A direct application of our method to continuous action spaces would not be advised, since it relies heavily on repeated visits to the same state (which will not naturally happen in continuous space).

\clearpage
\bibliographystyle{unsrt}
\bibliography{example_paper}

\clearpage
\section{Appendix}
\subsection{Factors influential to NN-CCE Performance}
In this section we provide a series of ablation studies to demonstrate the factors that affect the performance of our NN-CCE algorithm. Results are given comparing different versions of our agent against a fully random algorithm and MADDPG in small test scenarios within the multi particle environment. The specific scenario is ``simple" with 2 adversarial agents and 1 good agent. Results are given in terms of the total reward accumulated during a testing episode. Where either trained MADDPG or CCE agents control the adversarial players, and the other controls the good player. 

These results guided our development and implementation, contributing to the overall success for the larger tasks and test beds. 

\subsubsection{Number of trajectories}
Our agent uses significantly less environment interactions compared to the MADDPG algorithm. Because we do significant processing for each trajectory before neural network training, we restricted the number of trajectories sampled to at most half of that of MADDPG. 

We then trained MADDPG on a three small test scenarios in MPE and recorded its performance after 10k, 30k, and 50k trajectories through the environment, one for each test scenario. Our algorithm was then trained on the same small test scenarios but limiting the trajectories to 5k, 15k, and 25k trajectories respectively. 

For each small scenario, our algorithm scored at least $10\%$ better than MADDPG using at most half of the number of trajectories through the environment. For these scenarios we had NN-CCE play against MADDPG agents in a head-to-head however we recorded the average score 

In addition, the ratio of failure cases for MADDPG grew with the number of trajectories. We define a failure case as an instance where the policy post training does not improve significantly beyond the performance of a random policy. 

Interestingly, the ratio of failure cases for our agent decreased significantly as we increased the number of trajectories. Our agent makes a trade-off compared to other RL algorithms: we sample less trajectories from the environment but spend much longer processing the trajectories we do sample using no-regret learning. 

\subsubsection{Diversity of nodes within a layer}
Given the results from the previous section, we take significantly less trajectories through any environment we are training NN-CCE approximation in. As a result, we initially observed a wider range of performance for NN-CCE on the same environment over many random seeds. 

We discovered that NN-CCE agents that performed higher tended to have a higher spread in the value estimates for nodes across every layer. We measure spread for value's in each layer using the coefficient of variation (CV) for a given layer: $\frac{\sigma_h}{\mu_h}$.  

In order to utilize this observation as a reproducible process, we developed a subroutine within training that generates a fixed number of trees, measures the CV for each tree and a given layer, and uses the tree with highest CV for that layer. 

While this subroutine does increase the total number of environment trajectories, the agent still only trains on one of the trees generated. In addition this subroutine did not increase the maximum score in any of the test environments, instead it made the scores more consistent (less failure cases). 

\subsubsection{Strategic Dominance Action Pruning}
The goal of no-regret learning is to grow the regret with respect to the best action in hindsight sub-linearly. The algorithm will converge on to what it evaluates as the best action. If all players utilize no-regret learning, then their learned policies converge to the set of CCEs. 

Objectively speaking at a given state with N-players each having K strategies, there are multiple CCEs that exist, and our agent would converge to one of them. In fact we also found that successful MADDPG algorithms would converge onto one CCE for a given state as well. 

In classical game theory, having two or more competing equilibrium's for a state is not a problem as if equilibrium A was strictly better than equilibrium B, equilibrium B would not be an equilibrium by definition, but in MARL it can be an issue.

This is because the very definition of equilibrium assumes that all other players follow that recommended equilibrium. But in some cases of MARL, such as when our agent controls the 1 good agent, and MADDPG controls the 2 adversarial agents. Suddenly that assumption is violated, and the performance of our agent is due to random chance on how well our equilibrium compares to the opponent equilibrium; in scenarios like this, more than one player is not following the equilibrium we learned. 

Therefore it is important not only to learn how to play one equilibrium, but also learn to adapt to different equilibrium's for a given state. 

One way we found to improve performance around this problem is to prune the dominated strategies for all players before no-regret learning. Strategies that were deemed to be dominated by any other strategy were masked and not allowed for selection and their weights were ignored when converting weights to policy, thereby receiving a probability of selection of 0.  

We found that this optimization did not increase the maximum performance of our agent against MADDPG across any test scenario, but it did increase the mean performance against MADDPG by 23\% from an average score of 15.3 to 18.8 over many repeated test episodes and 10 random seeds. 

\subsubsection{Joint vs individual policy optimization}
Joint policy optimization has poor scaling to larger tasks but it allows for much higher express-ability of policies especially for competitive tasks, or tasks with competitive elements compared to optimizing solo policies. The problem is that in team based competitive tasks, if we view the player policies is random variables, the players should not be viewed as independent variables. Instead, they should be viewed as dependent variables (all players on the same team). A joint distribution created by multiplying the player marginal distributions cannot come close to the complexity of a joint distribution over all possible actions. It does not allow for coordination amongst the players and never will allow for such coordination. This is a problem for exploitability of a strategy in competitive settings. 

In cooperative tasks, this lack of express-ability isn't actually that much of a problem. Because we are just looking for the best joint action as a needle in the haystack, we don't need to consider All joint action futures individually, but can consider each players policy and form a joint distribution by multiplying the the marginals together (treating the players as independent). 

\subsubsection{Imbalanced data sets for equilibrium value prediction}
One of the highest impacting aspects of predicting unknown states' equilibrium values is the imbalance in value estimates accrued during simulation. This is particularly true in the first iteration where the policy sampling the states is effectively a random policy.

We can see here for the MPE environment allowing a random policy to sample leads to the following value distributions. It is heavily skewed in favor of the value 0, as it is very unlikely for random policies to collide leading to non-0 rewards for any given trajectory. 

When we attempted to train a Q-value network on this heavily biased data set, we found that, as expected, it primarily predicted a value of 0 for testing data; the errors for non-zero testing data was extremely high and variable. 

Therefore we applied up-sampling to minority values. Initially we separate the continuous value training data, (X,y) into K classes. Each class is defined as a non-overlapping range of size range(y)/K. In order to prevent over representation of a small set of data points during sampling, we maintain a rule that each class except one, $k_s$ must contain at least 1000 data points, where $k_s$ is defined as the class with the least data points. We therefore recursively combine the two smallest classes until the condition is met. 

\subsubsection{Trajectories}
There is one novel crucial component within the game tree creation algorithm: the partially random off-policy trajectories. In this component, a subset of players follow their policy but the rest are randomized. We found that a mix of on-policy and randomized learning agents improved the average performance of the final agent, compared to fully on-policy and fully random training, across multiple tasks within the MPE environment. 

For the purpose of score comparison between the three variations, we normalize the average and standard error of the fully random training performance to \textbf{$1 \pm 0.17$}, respectively. Fully on-policy training yielded an average performance of \textbf{$0.8 \pm 0.3$}, while partial randomization yielded an average performance of \textbf{$1.3 \pm 0.2$}. 

We attempted to use common place exploration vs. exploitation methods, such as UCB score from MCTS and epsilon learning from DQNs to improve performance, but neither significantly impacted performance in our simultaneous-move multi-agent setting. 

\subsubsection{NN supported policy and value estimation}
At the core of our algorithm is the estimation of non-stationary policies and values for any state and time pairing. For any given state, $s$, at time step $h$, we run a neural network supported bandit algorithm for a set number of iterations. During the bandit algorithm, we accumulate an average value for $s$, and after the bandit algorithm we obtain our policy for $s$. Once all states within time step $h$ have been processed, we train a NN, $N_h$, on the states of time-step $h$, and repeat the process for nodes in time-step $h-1$ using $N_h$ as our supplementary network. 

Compared to other MARL algorithms, such as MADDPG, PPO, or deep-MCTS, we trade-off a higher volume of data for attempting to get higher quality data. We can directly compare the effects of this layer by layer approach by comparing our results to those of stationary policy estimation on multiple tasks within the MPE environment. 

In stationary policy estimation, we no longer take a layer by layer approach, but instead accumulate value estimation by back propagating visited leaf nodes in the tree through the reverse trajectory used to reach them. Policy estimations, rather than using bandit algorithms at each state, are accumulated by visit count to each successor state. This approach is very similar to the approach in deep-MCTS. 

By normalizing the average and standard error of the stationary estimation performance to $1 \pm 0.56$, we find that our layer by layer method yields more robust and higher quality results with an average performance of $2 \pm 0.23$.

\subsubsection{Data processing and learning objectives}
After generating a game tree (Algorithm \ref{alg:GenGameTree}) we have a large data structure of states seen through environment interaction organized by time horizon. We will refer to ``layer h'' of our tree as as all nodes that are $h$ time steps away from any of the root nodes. 

Contrast to monte carlo tree search based algorithms, we do not attempt to approximate a value and policy for each state during the search itself, instead we process all nodes in reversed time order once the tree has been fully created.

SM multi-agent reinforcement learning opens the question as to how we should estimate both the value and policy of a given state. In classic RL and perfect information multi-agent RL, the value of a state is estimated using the bellman equation: $$V_{\pi}(s) = R(s) + \gamma*max_{a}V_{\pi}(s')$$ where $s' = T(s,a)$, and the policy can be determined by picking the action that leads to the next state with highest value for every state. 

The application of bellman equation to SM MARL becomes translucent as we would need to know the policy of all agents in order to make a value estimation. As mentioned previously, other MARL algorithms address this issue by using a non-learning policy for all other agents (MA-PPO), or keeping a local policy estimation for all other agents (MA-DDPG). Both of these options lead to potentially sub-optimal generalization as the performance of the agent is then directly tied to the opponent they trained with or against. 

In our study we begin by processing the nodes in layer $h=H$, the terminal nodes of the tree. Nodes in layer $H$ have the unique property that $V(s \in S_H) = R(s \in S_H)$ regardless of any policy, since the episode terminates after reaching state $S_H$. 

We begin by updating the associated weights for all nodes in layer $H$ where the loss of a given state, $s$, for player $n$ $l(s,n) = \frac{r_n(s)}{\max_{s \in S_H} r_n(s)}$ using Algorithm \ref{alg:MAEXP}, which returns a set of policy weights and state value estimates, $W_H, V_H$. 

As we are at $h=H$, we are only concerned with the value estimates, $V_H$. Combining these value estimates with the states of their parent nodes, $S_{H-1}$, we create a dataset for a Q-value estimation network of tuples ${(s,a,s',v)}$ where $s \in S_{H-1}$, $s' \in S_{H}$, $a$ is a valid action such that $T(s,a) \rightarrow s'$, and $v:= V(s') \in V_H$, and train a network on this data set. The process is then repeated until we reach $h=1$. 

\subsection{Simultaneous-Move MCTS}
We implemented our own form of Simultaneous-move MCTS based on the algorithm used in MuZero. 

It was adapted to fit simultaneous-move tasks where both teams/players picked a move simultaneously before it is sent to the environment. Value was calculated as an accumulated value for all iterations at a node, and policies were generated by child visit count. 

\begin{algorithm}[tb]
\caption{Generate Game Tree}
\label{alg:GenGameTree}
\textbf{Input}: $S$, State Space\\
\textbf{Input}: $\Delta$, Starting State Distribution \\
\textbf{Input}: $S_0 \subseteq S$, Set of starting states \\ 
\textbf{Input}: $A$, Action Space \\
\textbf{Input}: $H$, finite horizon \\
\textbf{Input}: $K$, number of simulations \\
\textbf{Input}: $T$, Transition Function \\
\textbf{Input}: $M$, Neural network model \\
\textbf{Output}: $R$, Set of root nodes 
\begin{algorithmic}[1] 
\State $R \gets \{\emptyset\} $ 
\For{$k = K, K -1 , \dots, 0$}
    \State $s_0 \gets \Delta(S_0)$
    \If{$s_0 \notin \{s' : N.state \in R\}$}
        \State $v_0, p_0 \gets$ Predict($M$, $s_0$)
        \State $N \gets Node(s_0, v_0, w \gets p_0)$
        \State $R \gets R \cup \{n\}$
    \EndIf
    \State $N \gets$ GetNode($R,s$)
    \State $s \gets s_0$, $w \gets p_0$, $\tau \gets 0$,
    \State $N.n \gets N.n + 1$
    \State $children \gets$ GetChildren($N$)
    \While{$\tau = 0$}
        \State Sample joint action, $j$, using $\frac{w_i}{\sum w_i}$ for all players
        \State $s' \gets T(s,j)$
        \If{$s' \notin children$}
            \State $v', p' \gets$ Predict(Model, $s'$)
            \State $N' \gets Node(s', v', w' \gets p')$
            \State $children \gets children \cup \{n'\}$
            \State $\tau = 1$
        \Else
            \State $s \gets s'$, $w \gets w'$
            \State $children \gets$ GetChildren($N'$) 
        \EndIf
        
    \EndWhile
\EndFor
\State return $R$
\end{algorithmic}
\end{algorithm}

\subsubsection{Data Storage and Sampling}
After the tree is generated we save every node to a replay buffer. Each node in the replay buffer is saved as a tuple defined as Transition(Node):
\begin{enumerate}
    \item $\{O_p\}$ for $p = 1...n$
    \item{Node value}
    \item{Node policy}
    \item{Node time step}
    \item{Node player}
    \item{Node Visit Count}
\end{enumerate}

Where $O_p$ is the observation for each player at Node. 
Because the goal of the NN is to replicate the tree value and policy estimation, we sample such that every node in the same time step has a uniform chance of selection. In addition we give each time step equal chance of selection. For instance if there is a task with 5 time steps, the overall probability would be divided to 20\% per time step. Within each time step, the 20\% is divided evenly amongst the nodes associated with that time step. Note that it is likely different time steps will have a different number of nodes. A further discussion of this method is given in the ablation studies. 

We also do not use priority sampling with our replay buffer. It is an extremely popular method that has been shown to drastically increase the speed of training (Schaul et al., 2015). However we believe it does not fit well with our data, and were not impressed with its early empirical results on training. 

Our data is atypical to other reinforcement learning algorithm. Typical RL algorithms store trajectory tuples of the form $(s,a,s',r)$, however we store information that has been aggregated over many iterations. Therefore the problem of rare experiences, that prioritized buffers address, is not as applicable. Sampling uniformly across layers provides better value and policy estimates in future iterations. 
\subsubsection{Machinery}
Neural networks and scenario simulations were performed on a local machine containing a NVIDIA GeForce RTX 2080 Ti graphics card (GPU) with 11gb of memory. In addition the machine contains a 3.70GHz Intel(R) CPU with 8 cores. 

\subsubsection{Neural Network Training}
Neural network architecture follows a standard DQN structure. It takes as input a batch of local observations and outputs a value estimate as well as policy estimate. There is a representation network that is malleable depending on the input type. If the observations are images then it is generated as a convolutional NN, otherwise it is a deep fully connected NN. The number of layers in the representation network are flexible to the performance needs, but we find 8 layers each of size 256 to be sufficient for all tasks in this study. 

After the data is passed through the representation network it is then passed through two separate networks, the value and policy network. The value network takes the representation output and produces a vector of variable size (support size) which is then converted to a scalar value. This method of estimating a value using a vector operation was popularized in Schrittwieser et al., 2019. It causes outputs to be between 0 and 1 and thus aids in numeric stability. The policy network also takes as input the representation output and produces a vector the size of the max action space of all agents. 

During training, the network predictions for each observation are measured against their stored node information counterparts stored in the replay buffer. We use cross entropy loss with stochastic gradient descent to optimize the network. Further implementation parameter details can be found in the appendix. 

The total procedure involves iterating between tree generation and neural network training. In order to measure intermediary progress, we measure the performance of our network against a random agent after each training session. We use the score against the random agent as both a validation procedure and termination procedure. If the agent scores worse against the random agent on an iteration, the training is undone and the iteration repeats itself. This is common procedure for iterating algorithms such as deep MCTS. If the network does not significantly improve its score after a series of iterations, the training process is terminated, and the last updated model is output.

\subsubsection{Sampling trajectories and generating a game tree}
We define a game tree is a series of Nodes, each representing a state, and directed edges, representing a transition between nodes. Algorithm \ref{alg:GenGameTree} provides a detailed account of how the game tree is generated and stored.

The algorithm returns a set of nodes, $R$, which contains the root node of $|R|$ trees generated in simulation. There is a unique tree for each unique starting state. If a task only has one unique starting state, then $|R| = 1$. 

Each node, $N$, stores the following information: $s$, state, $v$, value, $w$, policy weights, $children$, set of edges to child nodes, $parent$, edge to parent node

The simulation begins by choosing a root node from the starting distribution (line 3). If the state is new, we create a new root object and store it in our set of roots, $R$ (lines 4 - 7). If the state has been seen in a previous simulation, then we get the corresponding node object (line 9). In either case, we get the set of children and other variables (lines 10-12). 

The next step is to reach a leaf node. We define a leaf node as either a terminal node, or a node that does not yet exist in the tree. Upon reaching a leaf node, we create a new node and set the appropriate attributes (lines 16-19). 

\end{document}